# Measuring Marketing Communications Mix Effort Using Magnitude Of Influence And Influence Rank Metric

**Andry Alamsyah, Endang Sofyan, Tsana Hasti Nabila**
Faculty of Economics and Business, Telkom University
andrya@telkomuniversity.ac.id

**Abstract**

*In the context of modern marketing, Twitter is considered as a communication platform to spread information. Many companies create and acquire several Twitter accounts to support and perform varieties of marketing mix activities. Initially, each accounts used to capture specific market profile. Together, the accounts create network of information that provide consumer to the information they need depends on their contextual utilization. From many accounts available, we have the fundamental question on how to measure influence of each account in the market based not only their relations, but also the effects of their postings. Magnitude of Influence (MOI) metric is adapted together with Influence Rank (IR) measurement of accounts in their social network neighborhood. We use social network analysis approach to analyze 65 accounts in the social network of an Indonesian mobile phone network operator, Telkomsel which involved in marketing communications mix activities through series of related tweets. Using social network provide the idea of the activity in building and maintaining relationships with the target audience. This paper shows the results of the most potential accounts based on the network structure and engagement. Based on this research, the more number of followers one account has, the more responsibility it has to generate the interaction from their followers in order to achieve the expected effectiveness. The focus of this paper is to determine the most potential accounts in the application of marketing communications mix in Twitter.*

## 1. Introduction

The large amount of users on social media has turned *Twitter* into a communication platform to spread any kinds of information. *Twitter* also introduces a concept to assess the value of influence which can be obtained from the interaction as well as number of followers [1]. The various features offered by *Twitter*, are used by many companies to perform a variety of marketing activities.

Marketing communications is one of central activities in general marketing strategies, and *Twitter* is a cost-effective communication channel to apply marketing communications effort [2]. In the new communication environment, advertising is the central element of the marketing communications' programs, however, there are some other important elements to increase brand equity, as well as brand top of mind in consumer opinion. By performing various activities of marketing communications, it helps brand or product to build public brand awareness [3]. A method to establish the brand awareness is by disseminating product information in a social network [4]. Information dissemination process in a social network is described as a complex and repetitive information transfer process from one node to another in network, where often we get consumer attention after more than one transmission. This is the principle of social network propagation and dissemination behavior [5].

Each account in the social network represents the activities that are used by companies in building and maintaining their relationships with the target audience, which is referred as marketing communications mix [6]. To determine the potency and effectiveness of every account in the marketing communications mix activities, the influence value based on the network structure and the engagements which include number of followers, posts, as well as the interaction between nodes can be obtained. The measurement to obtain the value of influence based on the network structure and engagement is called the influence model which consists of Ratio of Affection (ROA), Magnitude of Influence (MOI), and Influence Rank (IR). All three metrics have relevance in the dissemination of information as well as various activities of marketing communications to determine the values of influence of each account [7].

Social Network Analysis (SNA) analyzes the social relationships in a network consisting of nodes and edges (also called the ties, link, or connection). Nodes represents the individual actors in the network, and edges represents relationships between the actors [8]. SNA borrows graph theory to model, visualize and quantify network [9]

In this paper, we use a case study from an Indonesian mobile phone network operator, *Telkomsel*, which is a market leader in the mobile telecommunications industry in Indonesia. By understanding how the social network works correctly, *Telkomsel* can build the brand awareness, attract new customers, conduct the brand intelligence activities and market research to maintain their position in the industry. Interacting with users on Twitter and other social media platforms is an obligatory activity to understand the degree of customer satisfaction, therefore, *Telkomsel* has been using various social media platforms such as Twitter and Facebook to maintain their relationship with the customer.

We identify 65 *Twitter* accounts related to the application of *Telkomsel* marketing communications mix which includes advertising, sales and promotions, and events and experiences of Telkomsel as the sample of the research. We classify the accounts into 6 different categories of various business fields, such as the *Telkomsel* main accounts, regional accounts, *Telkom* Group accounts, endorsers, communities, and partners. We apply MOI and IR metric to measure the influence of each account based on their relationships and their posts.

**2. Network And Engagement Metric**

Graph theory is used to quantify and visualize the relationships structure of individuals and other social actors, such as groups and organizations. It is also an approach to measure the importance or a popularity of a node in a network, which is called key players or centrality [10][11]. Based on a study [7], MOI and IR metric are used to quantify an account influence in the their network and their engagement with their followers.

*2.1. Magnitude of Influence*

Magnitude of Influence (MOI) is a metric that indicates the total influence of an account with selected posts and tweets in the social network. The value shows the total impression one account makes on a social network based not only on the followers, but also based on the whole tweet posted by a user [7].

To obtain the value of MOI, we use the principle function of *LCRT* which determines the interaction by using the number of favorites, mention, and retweet to identify the interaction or engagement of tweets. Each interaction mentioned is worth as 1, and if the tweet does not have any interaction it is worth 0. The function is defined as:

$$LCRT(v, p) = \begin{cases} 1 & if\ (v \in L(p) \cup C(p) \cup RT(p)) \\ 0 & Otherwise \end{cases} \quad (1)$$

where:
- *L(p)* is a function returning the number of people who favorite the tweet.
- *C(p)* is a function returning the number of people who mention the observing account in their tweet.
- *RT(p)*: is a function returning the number of people who retweet the tweet

The following definition is necessary in order to obtain the value of influence within a social network based on the metrics used after calculating the interaction with *LCRT* function:

- *F(v):* The function is symbolizing the number of people following a certain account *(v)* within a social network.
- *P(v):* The function is symbolizing the amount of tweets *(p)* posted by an account *(v)* in a social network.

We introduce Ratio of Affection (ROA) metric to describe tweets impression value of an account toward its followers as total of LCRT functions divided with the number of follower an account has.

$$ROA(v,p) = \frac{\sum_{v \in F(v)} LCRT(v,p)}{|F(v)|} \quad (2)$$

MOI metric is The root mean square of ROA metric compared to all posting posted by an account.

$$MOI(v) = \sqrt{\frac{\sum_{p \in P(v)} (ROA(v,p))^2}{|P(v)|}} \quad (3)$$

MOI is an essential metric because it shows the value of influence based on the proportions of followers and all tweets posted as a whole.

*2.2. Influence Rank*

Influence Rank (IR) quantify an influence based on the how well an account connected to others well connected account. This metric determines the most influential user by measuring each user that has relationships with other users which also has an important influence in a social network. A user with highest IR value is called the opinion leader. IR can be approximated by PageRank algorithm [7]. The PageRank algorithm is well known concept in centrality measurement in SNA [10][11]. The IR metric formulation is as follows:

$$PR(u) = \sum_{v \in B_u} \frac{PR(v)}{L(v)} \quad (4)$$

In the formula above, the *PR* value of a node *(u)* depends on the number of $B_u$ which is the total relationship that belongs to the node *(v)* with other nodes, and it is divided by the number of links or edge of a node or *L(v)*.

**3. Experiment, Result, and Analysis**

In our experiment, we observe 65 Telkomsel related account that divided into 6 account categories as we mention in part l, such as the *Telkomsel* main accounts, regional accounts, *Telkom* Group accounts, endorsers, communities, and partners. As described in Section II, we use MOI and IR metric to quantify each account influence over the overall *Telkomsel* social network. We collect twitter data until august 2015.

For IR metric calculations we construct network structure from the relations between 65 accounts. The edge represents undirected using friendship (follower-followed) relationship. For MOI metric calculations, we check each account and rank them according to each category. We split this chapter to network structure for IR calculations, and engagement that represented by MOI calculations.

## 3.1. Network Structure

Information disseminate to the receiver according to network structure. The path, density, diameter, cluster and other features of a networks influence how information propagate through a social network. For that reason, we construct the underlying network interaction between 65 accounts and calculate who is the most influential accounts according to their location in the network.

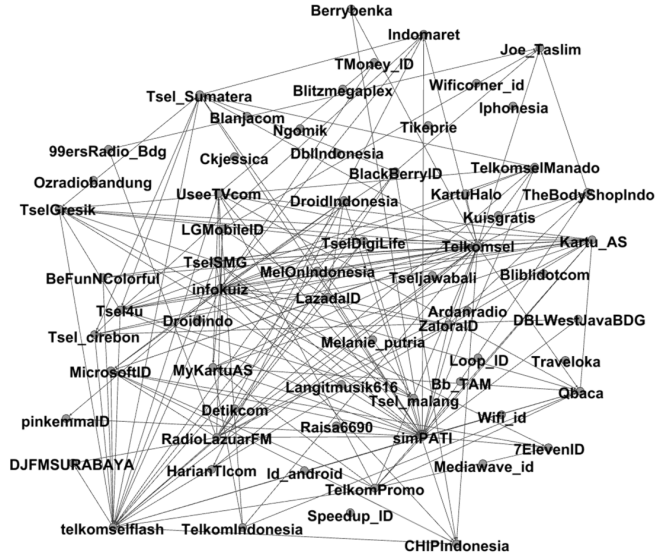

Fig. 1. The Telkomsel Social Network consists of 65 observed accouts

The network structure in Fig.1 is used to obtain the Influence Rank (IR) value. After implement the page rank algorithm calculations, we show 5 users with the highest Influence Rank (IR) value in the *Telkomsel* social network in Fig. 2. The user with the highest IR value is the opinion leader.

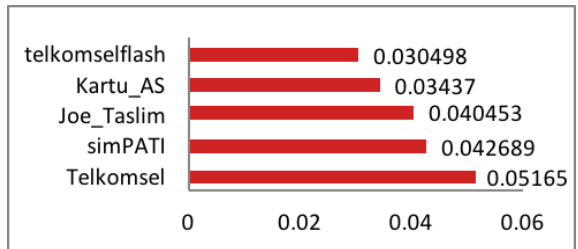

Fig. 2. Top five accounts with highest IR value in the social network

Based on the graph above, we learn that *Telkomsel* account is the most influential user, which means it has the most relationships with other users that also have impactful influence in the social network. In other words, *Telkomsel* is the opinion leader of the studied social network. This enables the account to spread the information wider than others account in the social network to build the brand awareness.

## 3.2. Engagement

We calculate MOI metric on each of 6 available categories. These following chart below show the MOI value of top 5 accounts on each category.

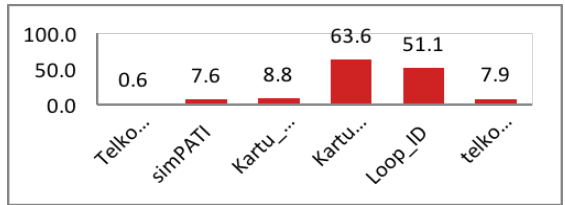

Fig. 3. The MOI values of the main users

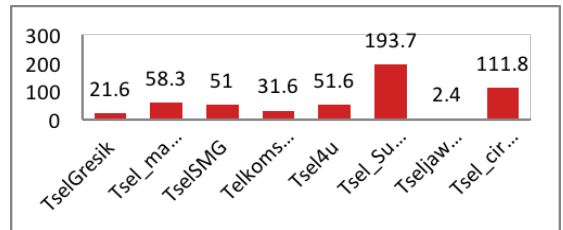

Fig. 4. The MOI values of the main users

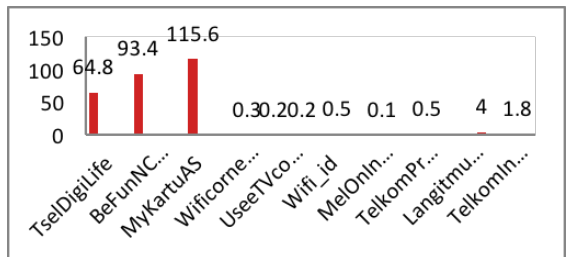

Fig. 5. The MOI values of the main users

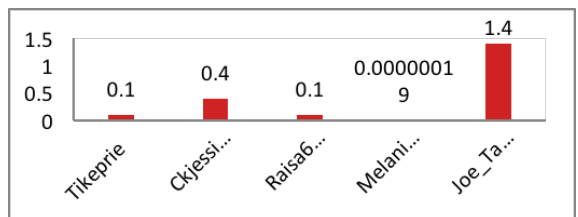

Fig. 6. The MOI values of the main users

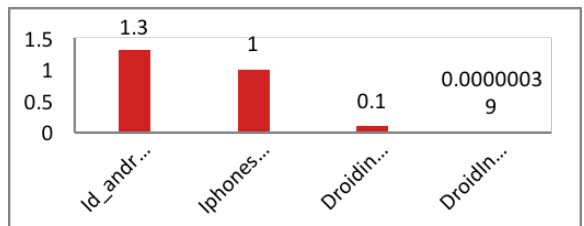

Fig. 7. The MOI values of the main users

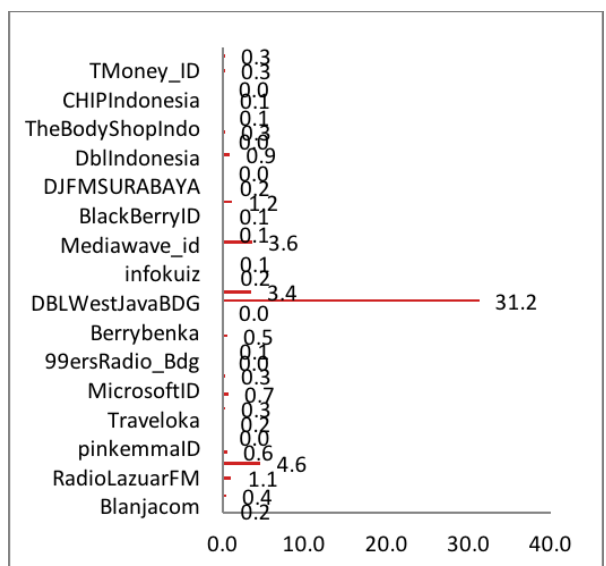

Fig. 8. The MOI values of the main users

Each account with highest MOI value as seen above shows their potential value in social media engagement compared with other similar accounts, as indicated in the same category. *KartuHalo* has the highest MOI for the main users with value of 63.6, *Tsel_Sumatera* has the highest MOI for the regional users with the value of 193.7, *MyKartuAS* has the highest MOI for Telkom Group with the value of 115.6, *Joe_Taslim* has the highest MOI for endorsers with the value of 1.4, *Id_android* has the highest MOI for communities with the value of 1.3, and *DBLWestJavaBDG* has the highest MOI of partners with the value of 31.2. Social media engagement value shows the capabilities of each account on how well they communicate the brand to get the attention of the audience. It also signifies that the highest value also shows the effectiveness of an account and how they present their content in social networks to build relationships with the target audience.

## 4. Conclusion

While Telkomsel is the opinion leader of the social network, its MOI value does not meet the expectation as a user with the highest value of IR. This means Telkomsel has an exceptional network to disseminate the information but not enough content that can engage wider user's interest. This also leads to a new finding that through the used metrics, the more number of followers one user has, the more responsibility they have to generate the interaction or engagement from their followers in order to achieve the the expected effectiveness.

To achieve the expected effectiveness, we suggest activities such as: 1) composing an interactive strategy to motivate its followers to participate in various contents; 2) creating richer and more appealing topics to reach the interest of followers; 3) comparing each account influence value for a targeted activity to reduce promotional cost.

This method can be proposed as a form of assessment in the selection of a product or brand influencer from *Twitter* and as an alternative for companies in measuring the effectiveness of the presentation done by the influencers through the social media.

For further research, we propose to analyze another object from different social media platform to obtain wider idea of the implementation of the metrics to various perspectives in business.